\documentclass[journal]{IEEEtran}
\IEEEoverridecommandlockouts

\usepackage{cite}
\usepackage{amsmath,amssymb,amsfonts, amsthm, bm}
\usepackage{algorithmic, algorithm}
\usepackage{graphicx, subcaption}
\usepackage{textcomp}
\usepackage{xcolor}
\usepackage{lipsum, hyperref, booktabs}

\def\BibTeX{{\rm B\kern-.05em{\sc i\kern-.025em b}\kern-.08em
    T\kern-.1667em\lower.7ex\hbox{E}\kern-.125emX}}

\begin{document}

\title{Deterministic Smoothed MAP Detection for High-Dimensional MIMO Systems
\thanks{This research work was supported by the French government, in the framework of BPI France 2030 program (INTENTION-6G project).}
}
\author{Vitor Tucci Ramos, Stephane Senecal, and Sheng~Yang,~\IEEEmembership{Member,~IEEE}
\thanks{V. Tucci Ramos and S. Senecal are with Orange, Orange Innovation, 92320 Ch\^atillon, France. S. Yang is with the Laboratoire des Signaux et Systèmes (L2S) at CentraleSup\'elec-CNRS-Universit\'e Paris-Saclay, 91192 Gif-sur-Yvette, France. Email:\{vitor.tucciramos, stephane.senecal\}@orange.com, sheng.yang@centralesupelec.fr }
}

\maketitle

\begin{abstract}

We propose a deterministic smoothed maximum a posteriori~(SMAP) framework for high-dimensional multiple-input multiple-output~(MIMO) detection. The discrete constellation prior is approximated by an i.i.d.\ Gaussian mixture, yielding a differentiable MAP objective that retains the constellation structure while enabling continuous optimization. Unlike sampling-based approaches, SMAP formulates detection as a deterministic optimization problem and can be initialized by the output of an arbitrary detector, allowing it to operate either as a standalone detector or as a refinement stage. For hard detection, we introduce SMAP-KR, which complements the continuous solution with a local $K_R$-neighbor search evaluated according to the original maximum-likelihood metric. Annealed continuation improves robustness for higher-order constellations, while a two-neighbor approximation reduces the cost of evaluating the mixture prior. We further develop a soft-output SMAP method in which a local Gaussian approximation based on the curvature of the smoothed posterior provides symbol probabilities and bit log-likelihood ratios without posterior sampling or explicit counterhypothesis lists. Numerical results for critically loaded MU-MIMO systems show that SMAP-KR provides an increasingly favorable performance--complexity tradeoff as the system dimension grows and can also effectively refine the output of existing detectors. For coded transmission, soft SMAP provides substantial block-error-rate gains over both LMMSE- and K-best-based soft detection.

\end{abstract}

\begin{IEEEkeywords}
MU-MIMO detection, integer least squares, maximum a posteriori estimation, smooth optimization, Gaussian mixture, soft-output detection.
\end{IEEEkeywords}
%


\section{Introduction}
\label{sec:introduction}

Recent advances in multiple-input multiple-output (MIMO) communications have led to the adoption of large antenna arrays at base stations (BSs) to improve spectral efficiency~\cite{MaMIMO,MaMIMO_survey,50years}. In the uplink, the BS must jointly detect the symbols transmitted by multiple user equipment~(UE) devices. As the number of simultaneously transmitted streams increases, this joint detection problem can become a major computational challenge.

Multiuser MIMO (MU-MIMO) detection can be formulated as a bounded integer least-squares~(ILS) problem, in which an integer-valued vector is estimated from a noisy linear observation through a known channel matrix. Under additive white Gaussian noise~(AWGN), maximum-likelihood (ML) detection is optimal but requires a combinatorial search over the discrete constellation. Bounded ILS detection is NP-hard in general, and the number of candidate vectors grows exponentially with both the detection dimension and the constellation size~\cite{ILS_comp}.

Importantly, high dimension alone does not necessarily imply difficult detection. When the number of receive antennas substantially exceeds the number of transmitted streams, favorable propagation can make the user channels nearly orthogonal, allowing simple linear detectors such as zero forcing~(ZF) and linear minimum mean-square error~(LMMSE) to perform well. Detection becomes substantially more challenging as the system load increases and the number of transmitted streams approaches the number of receive antennas. In this regime, the favorable geometry underlying linear detection is weakened, while near-ML discrete search remains prohibitively expensive in high dimensions.

High system loading is nevertheless attractive because it allows more data streams to share the available spatial resources. The challenge addressed in this work is therefore not high dimension alone, but high-dimensional detection at high system load. We focus on the critically loaded case, where the numbers of transmitted and received spatial dimensions are equal.

A broad class of MIMO detectors addresses this difficulty through structured discrete search~\cite{50years}. The classical sphere decoder~(SD) can achieve ML performance but has exponential worst-case complexity~\cite{MLD}. Fixed-width or fixed-complexity variants, such as the K-best sphere decoder~(K-SD)~\cite{K_best} and fixed-complexity sphere decoding~\cite{IFSD}, restrict the number of explored candidates to control computational cost. Such methods can provide excellent performance at moderate dimensions, but a fixed candidate budget becomes increasingly restrictive as the dimension and constellation order grow.

Lattice-reduction-aided detection provides another approach by transforming the channel basis into a more nearly orthogonal one before applying a linear or successive detector \cite{Yao_Wornell_LR,windpassinger2003low}. This can substantially improve the performance of low-complexity detectors without requiring exhaustive search. However, the basis-reduction preprocessing becomes increasingly costly in high dimensions, and conventional LR formulations are naturally suited to full-column-rank determined or underloaded systems. Overloaded systems require specialized extensions, typically involving additional candidate-search mechanisms~\cite{Hayakawa_LR_overloaded}. These considerations make LR-aided detection less attractive in the high-dimensional critically loaded or overloaded regimes of interest here.

An alternative is to relax the discrete constraint and solve a continuous surrogate problem. Linear detectors such as ZF and LMMSE have low complexity but can exhibit substantial performance loss in critically loaded or ill-conditioned systems. Semidefinite relaxation~(SDR) provides a tighter continuous approximation at increased computational cost. More generally, conventional continuous relaxations tend to weaken or discard the constellation structure during optimization, recovering it only when the continuous solution is mapped back to the discrete alphabet.

This has motivated methods that retain constellation information during the iterative detection procedure. Such information may be imposed implicitly through projections~\cite{GenPGD,PGD_net} or nonlinear denoising operations~\cite{ADMMNet,OAMP,OAMP_MA}. Other approaches introduce constellation-aware penalties or continuous representations directly into the optimization. For example, \cite{CMDNet} employed a Gumbel-softmax relaxation within a deep-unfolded detector, while \cite{SOAV_MAP} used a sum-of-absolute-values formulation and splitting methods. Homotopy and proximal formulations have also been considered~\cite{HOMTD}, and constellation-centered cosine and Gaussian penalties have been used in gradient-based detectors~\cite{GD_XL_MIMO,Q_penalty}. These approaches illustrate the benefit of preserving discrete-alphabet information while operating in a continuous domain.

A closely related probabilistic approach is to replace the discrete constellation prior by a smooth density. In particular, annealed Langevin dynamics~(ALD) detection~\cite{ALD} approximates the discrete prior by a Gaussian mixture and progressively decreases its variance.  The resulting smoothed posterior is explored through Langevin dynamics, combining deterministic score updates with injected Gaussian noise to generate candidate samples. Related stochastic-gradient Langevin approaches have also been considered~\cite{SGLD}, while other prior models have been investigated to improve exploration of the detection landscape~\cite{Tstu}. These methods demonstrate that smoothing the discrete prior provides a useful continuous representation of the constellation. Their primary use of the smoothed posterior, however, is as a distribution to be explored through stochastic dynamics.

In this work, we take a different view of the same underlying smoothing principle. Rather than sampling from the smoothed posterior, we treat its negative logarithm as an explicit deterministic optimization objective. We refer to the resulting framework as \emph{smoothed maximum a posteriori}~(SMAP) detection. This distinction has several consequences. The SMAP objective can be minimized using standard continuous optimization algorithms, and the optimization can be initialized by any available estimate. SMAP can therefore operate as a standalone detector or as a deterministic refinement stage applied to the output of another detector. Moreover, the continuous solution can be coupled to a controlled local search over the original discrete constellation, while the local curvature of the same objective can be used to construct soft information.

The main contributions of this work are summarized as follows:
\begin{itemize}
    \item We formulate a deterministic SMAP framework by replacing the i.i.d.~discrete constellation prior with an i.i.d.~Gaussian-mixture approximation and directly minimizing the resulting differentiable negative log-posterior. Unlike sampling-based use of the same smoothing principle, the resulting formulation is a deterministic optimization problem and can be initialized by the output of an arbitrary detector.
    \item For hard detection, we propose SMAP-KR, which complements the continuous SMAP solution with a $K_R$-neighbor discrete refinement.  A small set of constellation vectors closest to the continuous solution is generated efficiently and the final estimate is selected according to the original ML metric. We further introduce a two-neighbor approximation of the Gaussian-mixture prior to reduce its evaluation cost.
    \item For higher-order constellations, we develop an annealed SMAP strategy that progressively decreases the smoothing variance and warm-starts each optimization stage from the preceding solution, improving robustness to the increasingly multimodal objective.
    \item We develop a soft-output SMAP detector based on a local Gaussian approximation of the smoothed posterior. The Hessian of the SMAP objective provides coordinate-wise uncertainty estimates, from which symbol probabilities and bit log-likelihood ratios~(LLRs) are obtained without posterior sampling or explicit counterhypothesis lists.
    \item Numerical results for critically loaded MU-MIMO systems show that SMAP-KR becomes increasingly competitive as the detection dimension grows and can also effectively refine the output of an existing detector. For coded transmission, soft SMAP provides substantial block error rate (BLER) gains over both LMMSE- and K-best-based soft detection.
\end{itemize}

The remainder of the paper is organized as follows. Section~\ref{sec:sys} introduces the system model and the MAP detection problem. Section~\ref{sec:smap} develops the deterministic SMAP framework, including the prior approximation, continuous optimization, $K_R$-neighbor refinement, annealing, and their computational costs. Section~\ref{sec:soft_smap} develops the soft-output extension based on the local curvature of the SMAP objective. Numerical results, including hard- and soft-output performance and empirical computational cost, are presented in Section~\ref{sec:NR}. Finally, Section~\ref{sec:conclusion} concludes the paper.


\section{System Model and Problem Formulation}
\label{sec:sys}

We consider an uplink MU-MIMO system in which a base station equipped with $N_r$ receive antennas jointly detects the symbols transmitted by $N_t$ single-antenna users. The complex-valued received signal is modeled as
\begin{equation}
    \tilde{\pmb{y}} = \tilde{\pmb{H}}\tilde{\pmb{x}} + \tilde{\pmb{z}}, 
    \qquad \tilde{\pmb{z}} \sim \mathcal{CN}\!\left( \pmb{0}, \sigma_c^2\pmb{I} \right),
    \label{eq:sys_complex}
\end{equation}
where $\tilde{\pmb{y}}\in\mathbb{C}^{N_r}$ is the received vector, $\tilde{\pmb{H}}\in\mathbb{C}^{N_r\times N_t}$ is the channel matrix known to the receiver, $\tilde{\pmb{x}}\in\mathbb{C}^{N_t}$ is the transmitted symbol vector, and $\tilde{\pmb{z}}$ is circularly symmetric complex Gaussian noise. We focus on critically loaded systems, where $ N_t = N_r$. In this regime, detection can be particularly challenging when the channel matrix is ill-conditioned.

For the subsequent development, we use the equivalent real-valued representation
\begin{equation}
    \pmb{y} = \pmb{H}\pmb{x} + \pmb{z}, \label{eq:real_system_model}
\end{equation}
where
\begin{equation}
\begin{aligned}
\pmb{y} &=
\begin{bmatrix}
\Re\{\tilde{\pmb{y}}\}\\ \Im\{\tilde{\pmb{y}}\}
\end{bmatrix},
\
\pmb{x} =
\begin{bmatrix}
\Re\{\tilde{\pmb{x}}\}\\
\Im\{\tilde{\pmb{x}}\}
\end{bmatrix},
\
\pmb{z} =
\begin{bmatrix}
\Re\{\tilde{\pmb{z}}\}\\
\Im\{\tilde{\pmb{z}}\}
\end{bmatrix},
\\
\pmb{H} &=
\begin{bmatrix}
\Re\{\tilde{\pmb{H}}\} & -\Im\{\tilde{\pmb{H}}\}\\
\Im\{\tilde{\pmb{H}}\} &  \Re\{\tilde{\pmb{H}}\}
\end{bmatrix}.
\end{aligned}
\label{eq:real_equiv_model}
\end{equation}
Thus, $\pmb{H}\in\mathbb{R}^{m\times n}$, $\pmb{y}\in\mathbb{R}^{m}$, $\pmb{x}\in\mathbb{R}^{n}$, and $\pmb{z}\in\mathbb{R}^{m}$, with $m=2N_r$ and $n=2N_t$. Under the convention in \eqref{eq:sys_complex}, the real-valued noise satisfies
\begin{equation}
    \pmb{z} \sim \mathcal{N}\!\left( \pmb{0}, \sigma_z^2\pmb{I} \right), \qquad \sigma_z^2 \triangleq \frac{\sigma_c^2}{2}.
    \label{eq:real_noise}
\end{equation}
In what follows, $\sigma_z^2$ denotes the noise variance per real-valued dimension.

The transmitted vector belongs to the Cartesian product of the one-dimensional modulation alphabet $\mathcal{X}$, i.e., $\pmb{x}\in\mathcal{C}$ and $\mathcal{C} \triangleq \mathcal{X}^{n}$.  For square QAM constellations, $\mathcal{X}$ contains the possible in-phase or quadrature components of the complex symbols. For example, the real-valued representation of QPSK uses $\mathcal{X}=\{-1,+1\}$, whereas 16-QAM uses $\mathcal{X}=\{-3,-1,+1,+3\}$, up to the adopted constellation normalization. The transmitted coordinates are assumed to be independent and uniformly distributed over $\mathcal{X}$, so that
\begin{equation}
    p(\pmb{x}) = \prod_{i=1}^{n}p(x_i) = \frac{1}{|\mathcal{X}|^{n}}, \qquad
    \pmb{x}\in\mathcal{C}.
    \label{eq:discrete_uniform_prior}
\end{equation}

Given $\pmb{y}$ and $\pmb{H}$, maximum-likelihood~(ML) detection is defined as
\begin{equation}
    \hat{\pmb{x}}_{\mathrm{ML}} := \arg\max_{\pmb{x}\in\mathcal{C}} p(\pmb{y}\mid\pmb{x}).
    \label{eq:ml_prob}
\end{equation}
From \eqref{eq:real_noise}, the negative log-likelihood, up to terms independent of $\pmb{x}$, is
\begin{equation}
    -\log p(\pmb{y}\mid\pmb{x}) = \frac{1}{2\sigma_z^2} \left\| \pmb{y}-\pmb{H}\pmb{x} \right\|^2 +\mathrm{const.}
    \label{eq:negative_log_likelihood}
\end{equation}
Hence, ML detection reduces to the bounded discrete least-squares problem
\begin{equation}
    \hat{\pmb{x}}_{\mathrm{ML}} = \arg\min_{\pmb{x}\in\mathcal{C}} \left\| \pmb{y}-\pmb{H}\pmb{x} \right\|^2.
    \label{eq:ml_detection}
\end{equation}

Equivalently, the detection problem can be expressed in maximum a posteriori~(MAP) form as
\begin{align}
    \hat{\pmb{x}}_{\mathrm{MAP}} &=
    \arg\max_{\pmb{x}\in\mathcal{C}} p(\pmb{x}\mid\pmb{y}) \nonumber\\
    &=
    \arg\min_{\pmb{x}\in\mathcal{C}} \left\{
        \frac{1}{2\sigma_z^2} \left\| \pmb{y}-\pmb{H}\pmb{x} \right\|^2 - \log p(\pmb{x})
    \right\}.
    \label{eq:map}
\end{align}
Under the i.i.d.~discrete uniform prior in \eqref{eq:discrete_uniform_prior}, $-\log p(\pmb{x})$ is constant over $\mathcal{C}$, and therefore the MAP and ML detectors coincide.

Although the prior has no effect on the minimizer when optimization is restricted to the discrete feasible set $\mathcal{C}$, it provides a natural probabilistic representation of the constellation constraint. This observation motivates the approach developed in the next section: we extend the i.i.d.~discrete uniform prior to an i.i.d.~continuous Gaussian-mixture prior centered at the constellation points. The resulting smooth approximation preserves the constellation structure while allowing the MAP criterion to be optimized over a continuous domain.

\section{Deterministic Smoothed MAP Detection}
\label{sec:smap}

The discrete MAP problem in Section~\ref{sec:sys} involves a search over $M^n$ feasible vectors, where $M=|\mathcal{X}|$. We replace the discrete constellation prior with a continuous Gaussian-mixture approximation and directly minimize the resulting negative log-posterior. We refer to this deterministic optimization framework as smoothed maximum a posteriori~(SMAP) detection.

SMAP can operate standalone or be initialized by the output of another detector. For hard detection, we refine its continuous solution by searching the $K_R$ nearest feasible vectors and selecting the one with the smallest original ML metric; we refer to this realization as SMAP-KR. For higher-order constellations, an annealed variant progressively reduces the smoothing variance. The formulation and computational cost of these components are developed below.

\subsection{Gaussian-Mixture Relaxation of the Discrete Prior}
\label{subsec:smap_prior}

Recall that the real-valued transmitted symbols are independently drawn from the alphabet $\mathcal{X} = \{a_1,\ldots,a_M\}$, with uniform discrete prior
\begin{equation}
    p(x_i) = \frac{1}{M} \sum_{a\in\mathcal{X}} \delta(x_i-a).
    \label{eq:discrete_prior}
\end{equation}
We replace each Dirac mass by a Gaussian kernel of variance $\tau^2$, leading to the smoothed prior
\begin{equation}
    p_\tau(x_i) = \frac{1}{M} \sum_{a\in\mathcal{X}} \phi_\tau(x_i-a),
    \label{eq:smoothed_prior}
\end{equation}
where
\begin{equation}
    \phi_\tau(u) := \frac{1}{\sqrt{2\pi\tau^2}} \exp\left( -\frac{u^2}{2\tau^2} \right).
\end{equation}
Under the i.i.d.~prior assumption,
\begin{equation}
    p_\tau(\pmb{x}) = \prod_{i=1}^{n}p_\tau(x_i).
    \label{eq:vector_smoothed_prior}
\end{equation}

The parameter $\tau^2$ controls the degree of smoothing. As $\tau^2\rightarrow 0$, the Gaussian mixture concentrates around the original constellation points and approaches the discrete prior in the distributional sense. Larger values of $\tau^2$ produce a smoother continuous relaxation with weaker separation between neighboring modes.

For constellations having different minimum spacings, it is convenient to normalize the smoothing variance relative to the constellation geometry. If $d_{\min}$ denotes the minimum spacing of the real-valued alphabet, we use
\begin{equation}
    \tau^2 = \bar{\tau}^{\,2}d_{\min}^2, \label{eq:scaled_prior}
\end{equation}
where $\bar{\tau}^{\,2}$ is the normalized smoothing parameter. This allows the same smoothing regime to be used across different constellations.

\subsection{Smoothed MAP Objective}
\label{subsec:smap_objective}

Combining the Gaussian likelihood with the smoothed prior gives
\begin{equation}
    p_\tau(\pmb{x}\mid\pmb{y}) \propto \exp\left( -\frac{\|\pmb{y}-\pmb{H}\pmb{x}\|^2} {2\sigma_z^2} \right)
    \prod_{i=1}^{n}p_\tau(x_i).
\end{equation}
The SMAP estimate is defined as
\begin{equation}
    \pmb{x}_{\tau}^{\star} := \arg\min_{\pmb{x}\in\mathcal{B}^n} f_\tau(\pmb{x}),
    \label{eq:smap_problem}
\end{equation}
where, up to constants independent of $\pmb{x}$,
\begin{equation}
\begin{split}
    f_\tau(\pmb{x}) :=& \frac{1}{2\sigma_z^2} \|\pmb{y}-\pmb{H}\pmb{x}\|^2 \\ &- \sum_{i=1}^{n} \log \left[ \sum_{a\in\mathcal{X}} \exp\left( -\frac{(x_i-a)^2}{2\tau^2} \right) \right].
\end{split}
    \label{eq:smap_objective}
\end{equation}
Here, $\mathcal{B}:=[a_{\min},a_{\max}]$ spans the real-valued alphabet
and prevents the optimizer from exploring outside the constellation
range.

For later use, define the normalized Gaussian-mixture weights
\begin{equation}
    w_{i,a}(\pmb{x}) := \frac{ \exp\left(-\frac{(x_i-a)^2}{2\tau^2}\right) }{ \sum_{b\in\mathcal{X}} \exp\left(-\frac{(x_i-b)^2}{2\tau^2}\right) },
    \label{eq:posterior_weights}
\end{equation}
and the corresponding mixture-weighted constellation mean
\begin{equation}
    \mu_\tau(x_i) := \sum_{a\in\mathcal{X}} a\,w_{i,a}(\pmb{x}).
    \label{eq:posterior_mean}
\end{equation}
The gradient of~\eqref{eq:smap_objective} is
\begin{equation}
    \nabla f_\tau(\pmb{x}) = \frac{1}{\sigma_z^2} \pmb{H}^{\top}(\pmb{H}\pmb{x}-\pmb{y}) + \frac{1}{\tau^2} \left[ \pmb{x}-\boldsymbol{\mu}_\tau(\pmb{x}) \right],
    \label{eq:smap_gradient}
\end{equation}
where
$\boldsymbol{\mu}_\tau(\pmb{x}) := [\mu_\tau(x_1),\ldots,\mu_\tau(x_n)]^{\top}$.  Thus, the prior contribution attracts each continuous coordinate toward a weighted mean of the constellation points, with its strength and locality controlled by $\tau^2$.

\begin{figure*}[t!]
    \centering
    \begin{subfigure}[t]{0.28\textwidth}
        \centering
        \includegraphics[width=\textwidth]{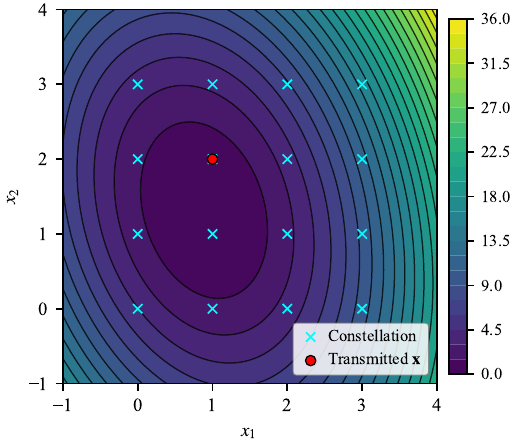}
        \caption{Likelihood term.}
        \label{fig:data_term}
    \end{subfigure}
    \hfill
    \begin{subfigure}[t]{0.28\textwidth}
        \centering
        \includegraphics[width=\textwidth]
        {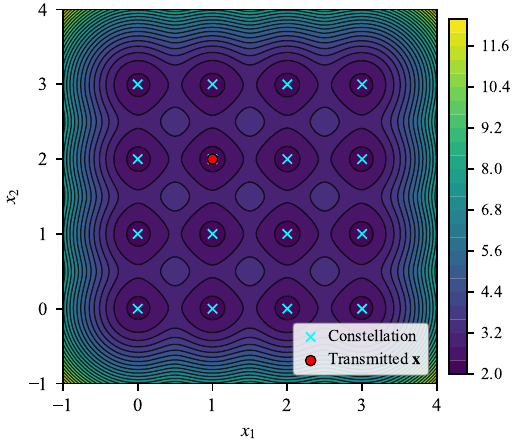}
        \caption{Smoothed prior term.}
        \label{fig:prior_term}
    \end{subfigure}
    \hfill
    \begin{subfigure}[t]{0.28\textwidth}
        \centering
        \includegraphics[width=\textwidth]
        {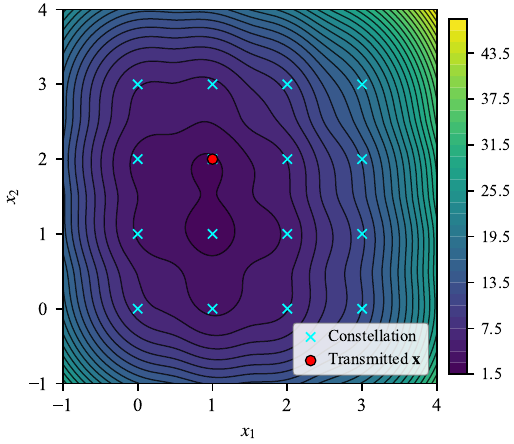}
        \caption{SMAP objective.}
        \label{fig:total_obj}
    \end{subfigure}
    \caption{Decomposition of the SMAP objective in a two-dimensional example ($\tau^2=0.13$, $\sigma_z^2=0.6$). Color indicates the value of each term.}
    \label{fig:prior_decomp}
\end{figure*}

The resulting landscape is illustrated in Fig.~\ref{fig:prior_decomp} for a two-dimensional example. The likelihood reflects the geometry induced by $\pmb{H}$, whereas the Gaussian-mixture prior introduces smooth modes centered at the feasible constellation vectors. Their combination produces a smooth but generally nonconvex objective shaped jointly by the observation and the constellation geometry.

For efficient implementation, define
\begin{equation}
    \pmb{A}:=\pmb{H}^{\top}\pmb{H},
    \qquad
    \pmb{v}:=\pmb{H}^{\top}\pmb{y}.
    \label{eq:precomputed_A_v}
\end{equation}
The gradient can then be written as
\begin{equation}
    \nabla f_\tau(\pmb{x}) = \frac{1}{\sigma_z^2} (\pmb{A}\pmb{x}-\pmb{v}) + \frac{1}{\tau^2} \left[ \pmb{x}-\boldsymbol{\mu}_\tau(\pmb{x}) \right].
    \label{eq:smap_gradient_precomputed}
\end{equation}
For $\pmb{H}\in\mathbb{R}^{m\times n}$, forming $\pmb{A}$ costs $\mathcal{O}(mn^2)$ once per channel realization, while forming $\pmb{v}$ costs $\mathcal{O}(mn)$ per received vector. Thereafter, the dense likelihood computation in each objective/gradient evaluation requires $\mathcal{O}(n^2)$ operations.

\subsection{Two-Neighbor Approximation}
\label{subsec:neighbor_approx}

For higher-order constellations, evaluating the Gaussian-mixture prior requires $\mathcal{O}(nM)$ operations per objective/gradient evaluation.  Since the Gaussian components decay exponentially with squared distance, we approximate the prior using only the two constellation points closest to each $x_i$. Denoting these points by $a_i^{(1)}$ and $a_i^{(2)}$, we use
\begin{equation}
    p_\tau(x_i) \simeq \frac{1}{M} \sum_{a\in\{a_i^{(1)},a_i^{(2)}\}} \phi_\tau(x_i-a),
    \label{eq:two_neighbor_prior}
\end{equation}
and compute the corresponding posterior mean from the same two components. The approximation is exact for binary alphabets.

For regularly spaced PAM alphabets, the two neighbors can be identified in constant time, reducing the prior-evaluation complexity to $\mathcal{O}(n)$. Hence, after precomputing $\pmb{A}$ and $\pmb{v}$, each SMAP objective/gradient evaluation is dominated by the dense matrix-vector product and has complexity $\mathcal{O}(n^2)$. The accuracy of this approximation is evaluated numerically in Section~\ref{subsec:params}.

\subsection{Deterministic Optimization and Initialization}
\label{subsec:smap_optimization}

We solve~\eqref{eq:smap_problem} using the limited-memory Broyden--Fletcher--Goldfarb--Shanno algorithm with box constraints (L-BFGS-B)~\cite{LBFGS_B}. This quasi-Newton method approximates the inverse Hessian using a limited history of gradient evaluations, thereby avoiding explicit Hessian formation or factorization during hard detection.

The default initialization is the continuous LMMSE estimate,
\begin{equation}
    \pmb{x}^{(0)} = \left( \pmb{H}^{\top}\pmb{H} + \lambda_{\mathrm{LMMSE}}\pmb{I} \right)^{-1} \pmb{H}^{\top}\pmb{y},
    \label{eq:lmmse_initialization}
\end{equation}
projected onto $\mathcal{B}^n$ when necessary.  More generally, however, $\pmb{x}^{(0)}$ can be obtained from any detector. In particular, a discrete estimate $\widehat{\pmb{x}}_0$ produced by a tree-search, message-passing, or other detection algorithm can be used directly as the initial point of the SMAP optimization. This property follows naturally from the deterministic optimization formulation and allows SMAP to operate either as a standalone detector or as a refinement stage for another algorithm.

Let $N_{\mathrm{fg}}$ denote the total number of objective/gradient evaluations performed during one L-BFGS-B optimization and let $N_{\mathrm{it}}$ denote the number of optimizer iterations. If $m_{\mathrm{LBFGS}}$ correction pairs are retained, the limited-memory vector operations require $\mathcal{O}(n m_{\mathrm{LBFGS}})$ per iteration. With the exact prior, the total continuous-optimization complexity is therefore $\mathcal{O}\Big( N_{\mathrm{fg}}(n^2+nM) + N_{\mathrm{it}}n m_{\mathrm{LBFGS}} \Big).$ With the two-neighbor approximation, this becomes
\begin{equation}
    \mathcal{O} \left( N_{\mathrm{fg}}n^2 + N_{\mathrm{it}}n m_{\mathrm{LBFGS}} \right).
    \label{eq:smap_optimization_complexity_TN}
\end{equation}

Since $m_{\mathrm{LBFGS}}$ is a small fixed memory parameter, the objective/gradient evaluations generally dominate. We use $N_{\mathrm{fg}}$ rather than only $N_{\mathrm{it}}$ in the complexity expression because the L-BFGS-B line search may evaluate the objective and gradient multiple times within one iteration. The actual $N_{\mathrm{fg}}$ is data dependent and is characterized empirically in Section~\ref{subsec:empirical_complexity}.


\subsection{Discrete \texorpdfstring{$K_R$}{KR}-Neighbor Refinement}
\label{subsec:KR_refinement}

The continuous SMAP solution $\pmb{x}_{\tau}^{\star}$ does not, in general, belong to $\mathcal{X}^n$. Component-wise quantization ignores the coupling introduced by $\pmb{H}$ and may therefore miss nearby constellation vectors with a smaller ML metric. We instead generate the $K_R$ feasible vectors closest to $\pmb{x}_{\tau}^{\star}$ in Euclidean distance and select
\begin{equation}
    \widehat{\pmb{x}}_{\mathrm{SMAP-KR}} := \arg\min_{\pmb{k}\in \mathcal{N}_{K_R}(\pmb{x}_{\tau}^{\star})} \|\pmb{y}-\pmb{H}\pmb{k}\|^2,
    \label{eq:KR_decision}
\end{equation}
where $\mathcal{N}_{K_R}(\pmb{x}_{\tau}^{\star})$ denotes the set of the $K_R$ nearest feasible vectors.

These vectors can be generated without enumerating $\mathcal{X}^n$.  For each coordinate, the alphabet points are ordered by their distance from $x_{\tau,i}^{\star}$. An index vector $\pmb{r}:=[r_1,\ldots,r_n]$ specifies a candidate, where $r_i$ is the rank of the selected symbol for coordinate $i$. Starting from $\pmb{r}=\pmb{1}$, candidates are generated by incrementing one index at a time and maintained in a min-heap according to their squared distance from $\pmb{x}_{\tau}^{\star}$, as summarized in Algorithm~\ref{alg:KClosest}.

\begin{algorithm}[t]
\caption{$K_R$ Closest Constellation Vectors}
\label{alg:KClosest}
\begin{algorithmic}[1]
\REQUIRE $\pmb{x}_{\tau}^{\star}$, $\mathcal{X}$, $K_R$
\ENSURE $\mathcal{N}_{K_R}(\pmb{x}_{\tau}^{\star})$

\FOR{$i=1,\ldots,n$}
    \STATE Order $\mathcal{X}$ by $|x_{\tau,i}^{\star}-a|^2$
\ENDFOR

\STATE Initialize $\pmb{r}=\pmb{1}$ and insert it into a
min-heap using its total squared distance as key
\STATE Initialize empty visited and candidate sets

\WHILE{fewer than $K_R$ candidates have been extracted}
    \STATE Extract the minimum-distance index vector $\pmb{r}$
    \IF{$\pmb{r}$ has not been visited}
        \STATE Add its constellation vector to $\mathcal{N}_{K_R}(\pmb{x}_{\tau}^{\star})$ and mark $\pmb{r}$ as visited
        \FOR{$i=1,\ldots,n$}
            \IF{$r_i<M$}
                \STATE Form $\pmb{r}'$ by setting $r_i'=r_i+1$
                \IF{$\pmb{r}'$ has not been visited}
                    \STATE Insert $\pmb{r}'$ into the heap
                \ENDIF
            \ENDIF
        \ENDFOR
    \ENDIF
\ENDWHILE
\end{algorithmic}
\end{algorithm}

Generic alphabet ordering costs $\mathcal{O}(nM\log M)$, while the regular PAM structure permits a more efficient implementation. Since each extracted candidate generates at most $n$ neighbors, generating $K_R$ candidates has the conservative complexity $\mathcal{O}\!\left(K_R n\log(K_Rn)\right).$ Evaluating their ML metrics costs $\mathcal{O}(K_Rmn)$ directly, or $\mathcal{O}(K_Rn^2)$ using the precomputed $\pmb{A}=\pmb{H}^{\top}\pmb{H}$ and $\pmb{v}=\pmb{H}^{\top}\pmb{y}$. Thus, the metric-evaluation cost can be written as $\mathcal{O}\!\left(K_R\min\{mn,n^2\}\right).$ The refinement therefore replaces exhaustive enumeration over $M^n$ vectors by a controlled local search whose cost is approximately linear in $K_R$ for the neighborhood sizes considered here.

\subsection{Annealed SMAP}
\label{subsec:annealing}

For small $\tau^2$, the smoothed prior has sharply separated modes, making the SMAP objective more sensitive to initialization, particularly for higher-order constellations. We therefore employ a deterministic continuation strategy with decreasing smoothing variances
\begin{equation}
    \tau_0^2 > \tau_1^2 > \cdots > \tau_{J-1}^2.
\end{equation}
At each stage $j$, L-BFGS-B is applied to $f_{\tau_j}(\pmb{x})$ over $\mathcal{B}^n$. The first stage uses the selected detector initialization, while each subsequent stage is warm-started from the solution of the preceding stage,
\begin{equation}
    \pmb{x}_{\mathrm{init}}^{(j)} = \pmb{x}^{(j-1)}, \qquad j=1,\ldots,J-1,
    \label{eq:annealing_stage}
\end{equation}
where $\pmb{x}^{(j)}$ denotes the locally converged solution returned at stage $j$. Thus, optimization begins with a smoother objective and progressively sharpens the constellation modes. The final solution $\pmb{x}^{(J-1)}$ is used as the continuous SMAP estimate and can be followed by the $K_R$-neighbor refinement of Section~\ref{subsec:KR_refinement}.

The complexity is the sum of the optimization costs across the $J$ stages. With the two-neighbor approximation, its dominant term is
\begin{equation}
    \mathcal{O}\!\left(
        n^2 \textstyle\sum_{j=0}^{J-1} N_{\mathrm{fg}}^{(j)} \right),
    \label{eq:annealed_complexity}
\end{equation}
where $N_{\mathrm{fg}}^{(j)}$ is the number of objective/gradient evaluations at stage $j$. Since each stage is warm-started, only a small number of continuation stages is needed in the considered experiments, as shown in Section~\ref{subsec:params}.

\subsection{Overall SMAP-KR Procedure}
\label{subsec:overall_smap}

Algorithm~\ref{alg:smap_kr} summarizes the complete hard-output procedure. Starting from an arbitrary initialization, SMAP is applied at one or more smoothing levels, followed by $K_R$-neighbor refinement using the original ML metric.

\begin{algorithm}[t]
\caption{SMAP-KR Detection}
\label{alg:smap_kr}
\begin{algorithmic}[1]
\REQUIRE $\pmb{y}$, $\pmb{H}$, $\mathcal{X}$, $\{\tau_j^2\}_{j=0}^{J-1}$, $\pmb{x}_{\mathrm{init}}$, $K_R$
\ENSURE Hard decision $\widehat{\pmb{x}}$

\STATE Compute or retrieve $\pmb{A}=\pmb{H}^{\top}\pmb{H}$ and $\pmb{v}=\pmb{H}^{\top}\pmb{y}$
\STATE Set $\pmb{x}^{(-1)}=\pmb{x}_{\mathrm{init}}$

\FOR{$j=0,\ldots,J-1$}
    \STATE Apply L-BFGS-B to $f_{\tau_j}(\pmb{x})$ over $\mathcal{B}^n$, initialized at $\pmb{x}^{(j-1)}$
    \STATE Denote the returned solution by $\pmb{x}^{(j)}$
\ENDFOR

\STATE Set $\pmb{x}_{\tau}^{\star}=\pmb{x}^{(J-1)}$
\STATE Generate $\mathcal{N}_{K_R}(\pmb{x}_{\tau}^{\star})$ using Algorithm~\ref{alg:KClosest}
\STATE Select $\displaystyle \widehat{\pmb{x}} = \arg\min_{\pmb{k}\in\mathcal{N}_{K_R}(\pmb{x}_{\tau}^{\star})} \|\pmb{y}-\pmb{H}\pmb{k}\|^2$

\RETURN $\widehat{\pmb{x}}$
\end{algorithmic}
\end{algorithm}

For a single smoothing stage with the two-neighbor approximation, the dominant per-vector complexity, excluding channel-dependent preprocessing, is
\begin{equation}
    \mathcal{O}\!\left( mn + N_{\mathrm{fg}}n^2 + K_R n\log(K_Rn) + K_R\min\{mn,n^2\} \right).
    \label{eq:overall_smap_complexity}
\end{equation}
Thus, SMAP-KR replaces exhaustive search over $M^n$ vectors by continuous optimization followed by a controlled local search. The resulting performance--complexity tradeoff is evaluated in Section~\ref{sec:NR}.

\section{Soft-Output SMAP Detection}
\label{sec:soft_smap}

Coded detection requires bit log-likelihood ratios (LLRs), whose exact computation involves prohibitive marginalization over the discrete posterior. We instead exploit the local curvature of the SMAP objective around the continuous solution to construct a local Gaussian approximation of the smoothed posterior. The resulting marginal means and variances are used to approximate symbol probabilities and bit LLRs.

\subsection{Local Gaussian Approximation}
\label{subsec:local_gaussian}

Recall that
\begin{equation}
    p_\tau(\pmb{x}\mid\pmb{y}) \propto \exp[-f_\tau(\pmb{x})].
\end{equation}
For soft-output generation, we assume that $\pmb{x}^{\star}$ is a locally converged stationary point of the unconstrained SMAP objective, so that
\begin{equation}
    \nabla f_\tau(\pmb{x}^{\star}) \simeq \pmb{0}.
\end{equation}
Such a point can be obtained, for example, by removing the box constraint and continuing the optimization from the SMAP solution used for hard detection.

A second-order expansion around $\pmb{x}^{\star}$ then gives
\begin{equation}
    f_\tau(\pmb{x}) \simeq f_\tau(\pmb{x}^{\star}) + \frac{1}{2} (\pmb{x}-\pmb{x}^{\star})^{\top} \pmb{J}^{\star} (\pmb{x}-\pmb{x}^{\star}),
    \label{eq:quadratic_local_approx}
\end{equation}
where
\begin{equation}
    \pmb{J}^{\star} := \nabla^2 f_\tau(\pmb{x}^{\star})
    \label{eq:local_precision}
\end{equation}
is the local curvature matrix.

When $\pmb{J}^{\star}$ is positive definite, this yields
\begin{equation}
    p_\tau(\pmb{x}\mid\pmb{y}) \approx \mathcal{N} \left( \pmb{x}^{\star}, (\pmb{J}^{\star})^{-1} \right).
    \label{eq:local_gaussian}
\end{equation}
This approximation captures the posterior locally around the selected SMAP solution and does not represent distant modes.

\subsection{Curvature of the Smoothed MAP Objective}
\label{subsec:smap_hessian}

Define the coordinate-wise variance of the constellation points under
the Gaussian-mixture weights in~\eqref{eq:posterior_weights} as
\begin{equation}
    \gamma_\tau(x_i) := \sum_{a\in\mathcal{X}} w_{i,a}(\pmb{x}) \left(a-\mu_\tau(x_i)\right)^2.
    \label{eq:posterior_variance}
\end{equation}
Differentiating~\eqref{eq:posterior_mean} gives
\begin{equation}
    \frac{d\mu_\tau(x_i)}{dx_i} = \frac{\gamma_\tau(x_i)}{\tau^2},
\end{equation}
and hence
\begin{equation}
    \frac{d^2[-\log p_\tau(x_i)]}{dx_i^2} = \frac{1}{\tau^2} - \frac{\gamma_\tau(x_i)}{\tau^4}.
    \label{eq:prior_curvature}
\end{equation}

Defining
\begin{equation}
    \boldsymbol{\Gamma}_{\tau}(\pmb{x}) := \operatorname{diag} \left( \gamma_\tau(x_1),\ldots,\gamma_\tau(x_n) \right),
\end{equation}
the Hessian of the SMAP objective is
\begin{equation}
    \nabla^2 f_\tau(\pmb{x}) = \frac{1}{\sigma_z^2}\pmb{H}^{\top}\pmb{H} + \frac{1}{\tau^2}\pmb{I} - \frac{1}{\tau^4} \boldsymbol{\Gamma}_{\tau}(\pmb{x}).
    \label{eq:smap_hessian}
\end{equation}
Thus, at the SMAP solution,
\begin{equation}
    \pmb{J}^{\star} = \frac{1}{\sigma_z^2}\pmb{A} + \frac{1}{\tau^2}\pmb{I} - \frac{1}{\tau^4} \boldsymbol{\Gamma}_{\tau}(\pmb{x}^{\star}),
    \label{eq:smap_hessian_solution}
\end{equation}
where $\pmb{A}=\pmb{H}^{\top}\pmb{H}$ is already available from hard detection. The prior curvature can be negative when $\gamma_\tau(x_i)>\tau^2$, so $\pmb{J}^{\star}$ is not necessarily positive definite.

Forming $\pmb{J}^{\star}$ costs $\mathcal{O}(n^2)$ once $\pmb{A}$ is available. Computing the coordinate-wise variances costs $\mathcal{O}(nM)$ with the exact prior and $\mathcal{O}(n)$ with the two-neighbor approximation.

\subsection{Curvature Regularization and Marginal Variances}
\label{subsec:curvature_regularization}

Since $\pmb{J}^{\star}$ is not necessarily positive definite, we regularize it before computing the marginal uncertainties:
\begin{equation}
    \widetilde{\pmb{J}}^{\star} := \pmb{J}^{\star}+\delta\pmb{I},
    \qquad
    \delta := \max\!\left\{ 0,\, \epsilon-\lambda_{\min}(\pmb{J}^{\star}) \right\},
    \label{eq:regularized_hessian}
\end{equation}
where $\epsilon>0$ is a small numerical floor. Thus, $\widetilde{\pmb{J}}^{\star}\succeq\epsilon\pmb{I}$, while no shift is applied when $\pmb{J}^{\star}$ is already sufficiently positive definite.

The marginal variances are obtained from the diagonal of the inverse curvature matrix,
\begin{equation}
    s_i^2 = \bigl[ (\widetilde{\pmb{J}}^{\star})^{-1} \bigr]_{ii}, \qquad i=1,\ldots,n,
    \label{eq:marginal_variance}
\end{equation}
yielding
\begin{equation}
    x_i\mid\pmb{y} \approx \mathcal{N}(x_i^{\star},s_i^2).
    \label{eq:marginal_gaussian}
\end{equation}

Computing all diagonal entries of $(\widetilde{\pmb{J}}^{\star})^{-1}$, e.g., through a Cholesky factorization of the dense curvature, has complexity $\mathcal{O}(n^3)$ and dominates the additional cost of soft-output generation. It is performed only once per continuous SMAP solution, after which the same marginal variances are used to compute all symbol probabilities and bit LLRs.

\subsection{Symbol Probabilities}
\label{subsec:symbol_probabilities}

To map the continuous Gaussian marginals to the discrete alphabet, let $\mathcal{X}:=\{a_1<\cdots<a_M\}$ and define the nearest-neighbor decision thresholds
\begin{equation}
  b_j:=\frac{a_j+a_{j+1}}{2},\quad j=1,\ldots,M-1,
\end{equation}
and $b_0:=-\infty,\quad b_M:=+\infty.$ Under the local Gaussian approximation, the probability assigned to symbol $a_j$ is
\begin{equation}
\begin{split}
    P_{i,j} &\triangleq \Pr(x_i\in[b_{j-1},b_j)\mid\pmb{y}) \\ &\approx \Phi\!\left( \frac{b_j-x_i^{\star}}{s_i} \right) - \Phi\!\left( \frac{b_{j-1}-x_i^{\star}}{s_i} \right),
\end{split}
    \label{eq:symbol_probability}
\end{equation}
where $\Phi(\cdot)$ is the standard Gaussian cumulative distribution function. These probabilities are automatically normalized, while $s_i^2$ controls the spread of probability among neighboring symbols.

Computing all symbol probabilities requires $\mathcal{O}(nM)$ operations, which is small compared with the $\mathcal{O}(n^3)$ curvature factorization.

\subsection{Bit LLRs}
\label{subsec:bit_llrs}

Let each constellation point carry $q=\log_2 M$ bits, and define
\begin{equation}
    \mathcal{X}_{\ell,b} := \{a\in\mathcal{X}:b_\ell(a)=b\}, \qquad b\in\{0,1\},
\end{equation}
where $b_\ell(a)$ denotes the $\ell$th bit of the label of $a$. Using
the symbol probabilities in~\eqref{eq:symbol_probability}, the
corresponding LLR is
\begin{equation}
    L_{i,\ell} = \log \frac{ \sum_{a_j\in\mathcal{X}_{\ell,0}} P_{i,j} }{ \sum_{a_j\in\mathcal{X}_{\ell,1}} P_{i,j} }.
    \label{eq:smap_llr}
\end{equation}
A small probability floor is used in implementation to avoid numerical overflow.

Computing all bit LLRs requires $\mathcal{O}(nM)$ operations. Unlike list-based soft detection, soft SMAP does not require explicit candidates for both hypotheses of each bit; reliability is instead obtained from the local posterior approximation around the continuous SMAP solution.

\subsection{Soft-Output SMAP Procedure}
\label{subsec:soft_smap_algorithm}

Algorithm~\ref{alg:soft_smap} summarizes the soft-output procedure.

\begin{algorithm}[t]
\caption{Soft-Output SMAP Detection}
\label{alg:soft_smap}
\begin{algorithmic}[1]
\REQUIRE $\pmb{x}^{\star}$, $\pmb{A}=\pmb{H}^{\top}\pmb{H}$, $\sigma_z^2$, $\tau^2$, $\mathcal{X}$, bit labeling
\ENSURE Bit LLRs $\{L_{i,\ell}\}$

\FOR{$i=1,\ldots,n$}
    \STATE Compute $\gamma_\tau(x_i^\star)$ from the Gaussian-mixture weights
\ENDFOR
\STATE Form $\pmb{J}^{\star}$ using \eqref{eq:smap_hessian_solution}
\STATE Regularize $\pmb{J}^{\star}$ using
\eqref{eq:regularized_hessian}
\STATE Compute the marginal variances $s_i^2$ using
\eqref{eq:marginal_variance}

\FOR{$i=1,\ldots,n$}
    \STATE Compute $\{P_{i,j}\}_{j=1}^{M}$ using
    \eqref{eq:symbol_probability}
    \STATE Compute $\{L_{i,\ell}\}_{\ell=1}^{\log_2 M}$ using
    \eqref{eq:smap_llr}
\ENDFOR

\RETURN $\{L_{i,\ell}\}$
\end{algorithmic}
\end{algorithm}

After obtaining the continuous SMAP solution, the additional soft-output complexity is
\begin{equation}
    \mathcal{O}(n^3+nM),
    \label{eq:soft_complexity}
\end{equation}
dominated by the dense curvature factorization. Thus, hard and soft SMAP share the same continuous optimization stage, with curvature estimation constituting the principal additional cost of soft detection.

The proposed soft output is a local approximation: it captures uncertainty around the selected SMAP solution but does not explicitly represent probability mass associated with distant posterior modes. Unlike sampling or list-based methods, however, it requires neither posterior sampling nor explicit counterhypothesis candidates, and derives the soft information directly from the same SMAP objective used for detection. The smoothing variance value used in Algorithm~\ref{alg:soft_smap} is the final scheduled annealing variance, $\tau_{J-1}^2$. Its coded performance is evaluated in Section~\ref{subsec:coded_perf}.

\section{Numerical Results}
\label{sec:NR}

We evaluate the hard- and soft-output SMAP detectors in critically loaded MU-MIMO systems. We first examine the main design parameters and then compare SMAP with representative linear, iterative, relaxation-based, and tree-search detectors. Empirical computational cost, system load and coded performance are also evaluated.

\subsection{Simulation Setup}
\label{subsec:sim_params}

We consider the critically loaded uplink system of Section~\ref{sec:sys}, with $N_r=N_t$ and real-valued dimension $n=2N_t$. The complex channel entries are independently distributed as
\begin{equation}
    [\widetilde{\pmb{H}}]_{ij} \sim \mathcal{CN}\!\left(0,\frac{1}{N_t}\right),
    \label{eq:simulation_channel}
\end{equation}
and perfect channel knowledge is assumed at the receiver. The real-valued alphabets are $\mathcal{X}=\{-1,+1\}$ for QPSK and $\mathcal{X}=\{-3,-1,+1,+3\}$ for 16-QAM. Higher-order alphabets are also considered for selected parameter studies.

The following reference detectors are considered:
\begin{itemize}
    \item \textbf{LMMSE:} linear MMSE detection, also used as the default initialization for standalone SMAP;
    \item \textbf{K-best SD:} fixed-width tree search retaining $K_{\mathrm{SD}}$ candidates per level \cite{K_best};
    \item \textbf{K-best SQRD:} K-best detection with sorted QR decomposition \cite{SQRD};
    \item \textbf{OAMP:} orthogonal approximate message passing with alternating linear estimation and nonlinear denoising \cite{OAMP};
    \item \textbf{SDR-RBR:} row-by-row semidefinite-relaxation detection \cite{SDR-RBR}.
\end{itemize}

Unless otherwise stated, $K_{\mathrm{SD}}=90$ for QPSK and $K_{\mathrm{SD}}=120$ for 16-QAM. OAMP uses a damping factor of $0.3$ and at most 60 iterations, while SDR-RBR uses at most 30 iterations and is considered only for binary real-valued alphabets. Uncoded results are averaged over 5000 independent channel realizations.

The SNR is defined as
\begin{equation}
    \mathrm{SNR} \triangleq \frac{ \mathbb{E}[\|\pmb{H}\pmb{x}\|_2^2] }{ \mathbb{E}[\|\pmb{z}\|_2^2] }.
    \label{eq:simulation_snr}
\end{equation}

\subsection{Performance Metrics and Empirical ML Lower Bound}
\label{subsec:perf_metrics}

For uncoded transmission, we report symbol error rate~(SER) and vector error rate~(VER), where a vector error occurs if at least one real-valued symbol is incorrectly detected. For coded transmission, performance is measured by block error rate~(BLER).

Since exact ML detection is computationally prohibitive at the dimensions considered here, we use the empirical ML error lower bound introduced in \cite{CY18}. For each realization, let
\begin{equation}
    d_{\mathrm{true}} := \|\pmb{y}-\pmb{H}\pmb{x}_{\mathrm{true}}\|^2
\end{equation}
and, for a pool $\mathcal{P}$ of feasible candidates generated by the considered detectors, define
\begin{equation}
    d_{\mathrm{pool}} := \min_{\pmb{k}\in\mathcal{P}} \|\pmb{y}-\pmb{H}\pmb{k}\|^2.
\end{equation}
Whenever
\begin{equation}
    d_{\mathrm{pool}}<d_{\mathrm{true}},
    \label{eq:certified_ml_error}
\end{equation}
an ML error is certain, since
\begin{equation}
    \|\pmb{y}-\pmb{H}\pmb{x}_{\mathrm{ML}}\|^2 \leq d_{\mathrm{pool}}<d_{\mathrm{true}}.
\end{equation}
Importantly, this certification does not require the pool to contain the ML solution; any feasible vector with a metric below that of the transmitted vector is sufficient.

Thus, denoting the certified event in \eqref{eq:certified_ml_error} by $\mathcal{E}_{\mathrm{cert}}$,
\begin{equation}
    \mathcal{E}_{\mathrm{cert}} \subseteq \mathcal{E}_{\mathrm{ML}},
    \qquad
    \Pr(\mathcal{E}_{\mathrm{cert}}) \leq \Pr(\mathcal{E}_{\mathrm{ML}}).
    \label{eq:ml_lower_bound}
\end{equation}
The empirical frequency of $\mathcal{E}_{\mathrm{cert}}$ is therefore reported as a lower bound on the ML error probability. Although the bound may be loose, it becomes tighter as the candidate pool improves. More importantly, it is inexpensive to evaluate even in very high dimensions and can provide a nontrivial lower bound on the fundamental ML error probability when exact ML detection is computationally infeasible.

\subsection{SMAP-KR Design Parameters}
\label{subsec:params}

We first examine the main SMAP-KR design parameters: the smoothing variance, initialization, two-neighbor prior approximation, refinement size, and annealing schedule. All smoothing variances are reported in normalized form, $\bar{\tau}^{\,2}=\tau^2/d_{\min}^2$.

\subsubsection{Smoothing Variance}

\begin{figure}[t]
    \centering
    \includegraphics[width=0.7\linewidth,keepaspectratio]
    {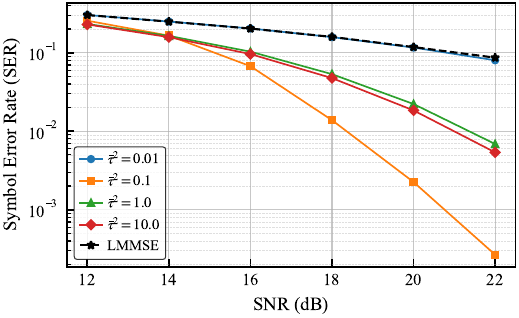}
    \caption{SER for different normalized smoothing variances $\bar \tau^2$
    ($n=80$, $|\mathcal{X}|=4$).}
    \label{fig:ser_vs_tau}
\end{figure}

Figure~\ref{fig:ser_vs_tau} illustrates the effect of $\bar \tau^2$. Small values preserve strong attraction toward the constellation but produce
a highly multimodal objective, whereas large values improve smoothness at the expense of weakening the prior. As $\bar \tau\rightarrow\infty$, the SMAP objective approaches box-constrained least squares. The best performance is therefore obtained at an intermediate smoothing level. Based on Fig.~\ref{fig:ser_vs_tau}, we chose $\bar\tau^2=0.1$ for single-stage detection and scale it according to~\eqref{eq:scaled_prior} for other spacings, unless otherwise stated.

\subsubsection{Initialization and Detector Refinement}

\begin{figure}[t]
    \centering
    \includegraphics[width=0.7\linewidth,keepaspectratio]
    {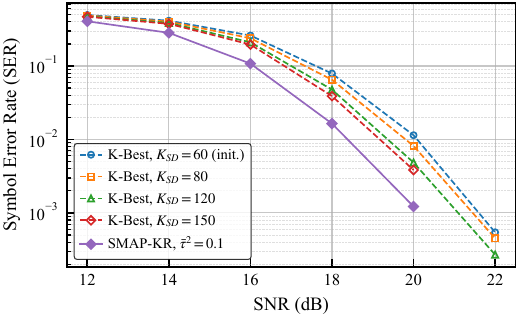}
    \caption{SER comparison for SMAP initialized by K-best SQRD with
    $K_{\mathrm{SD}}=60$ ($n=120$, $|\mathcal{X}|=8$).}
    \label{fig:ser_vs_init}
\end{figure}

A useful property of deterministic SMAP is that it can be initialized by any available detector. Figure~\ref{fig:ser_vs_init} considers K-best SQRD with $K_{\mathrm{SD}}=60$ as the initializer. SMAP refinement improves its output and outperforms standalone K-best SQRD with larger candidate widths. Thus, in this setting, additional computation is more effectively allocated to SMAP refinement than solely to enlarging the first-stage tree search.

\subsubsection{Two-Neighbor Prior Approximation}

\begin{figure}[t]
    \centering
    \includegraphics[width=0.7\linewidth,keepaspectratio]
    {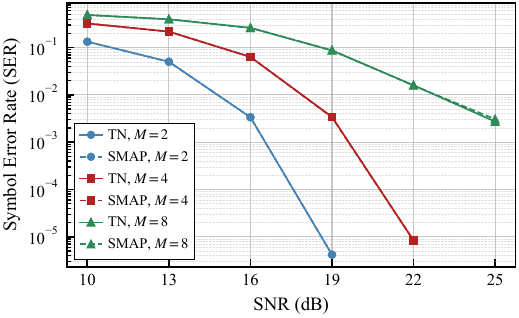}
    \caption{SER comparison between the exact prior and the two-neighbor approximation ($n=80$).}
    \label{fig:ser_TN_approx}
\end{figure}

Figure~\ref{fig:ser_TN_approx} compares the exact Gaussian-mixture prior with the two-neighbor approximation in \eqref{eq:two_neighbor_prior}. The two methods provide nearly identical SER for the evaluated higher-order alphabets, indicating that distant Gaussian components have negligible influence in the operating regime of interest. The approximation is exact for $|\mathcal{X}|=2$ and reduces the prior-evaluation complexity from $\mathcal{O}(nM)$ to $\mathcal{O}(n)$.

\subsubsection{Number of Refinement Candidates}

\begin{figure}[t!]
    \centering
    \includegraphics[width=0.7\linewidth]
    {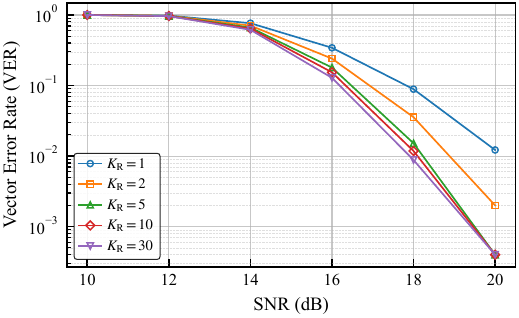}
    \caption{VER for different values of $K_R$ ($n=80$, $|\mathcal{X}|=2$).}
    \label{fig:ser_multiple_taus}
\end{figure}

\begin{table}[t!]
\centering
\setlength{\tabcolsep}{3pt}
\begin{tabular}{l c c c c c}
\toprule
\textbf{$K_R$} & 1 & 2 & 5 & 10 & 30 \\
\midrule
Running time (ms) & 0.33 & 0.47 & 1.01 & 1.84 & 5.79 \\
\bottomrule
\end{tabular}
\caption{Running time of the $K_R$-neighbor refinement for different neighborhood sizes.}
\label{tab:k_runtime}
\end{table}

As shown in Fig.~\ref{fig:ser_multiple_taus}, increasing $K_R$ improves performance by allowing nearby feasible vectors with a smaller ML metric to be recovered, although the gain eventually saturates. Table~\ref{tab:k_runtime} shows that the measured refinement time grows approximately linearly with $K_R$ over the considered range. Based on this performance--complexity tradeoff, we use $K_R=5$ for the subsequent binary-alphabet experiments.

\subsubsection{Annealing}

\begin{figure}[t!]
    \centering
    \includegraphics[width=0.7\linewidth,keepaspectratio]
    {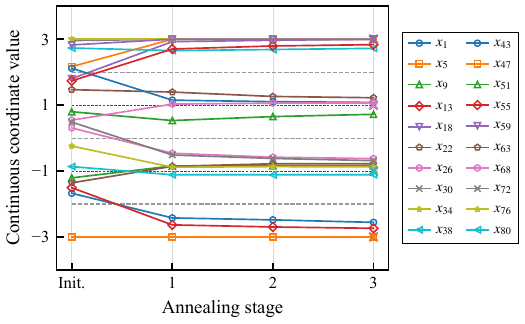}
    \caption{Evolution of representative continuous symbol estimates during SMAP annealing.}
    \label{fig:annealing_trace}
\end{figure}

\begin{figure}[t!]
    \centering
    \includegraphics[width=0.7\linewidth]
    {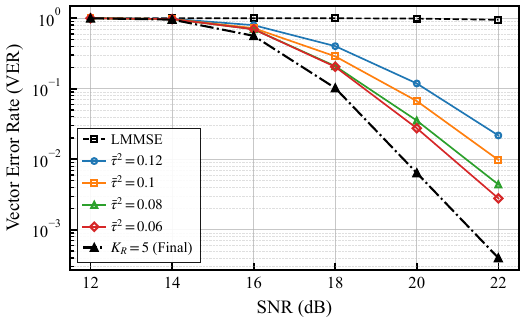}
    \caption{VER after successive annealing stages ($n=80$, $|\mathcal{X}|=4$).}
    \label{fig:ser_intermediate_taus_07}
\end{figure}

For higher-order constellations, annealing progressively reduces the smoothing variance while warm-starting each stage from the preceding solution. Figure~\ref{fig:annealing_trace} illustrates this process at the coordinate level. Several estimates cross decision boundaries as the prior modes sharpen, showing how continuation can correct decisions that would result from quantizing an earlier-stage solution.

Figure~\ref{fig:ser_intermediate_taus_07} shows the corresponding performance after successive stages. Most of the improvement is obtained in the early stages, while subsequent stages provide diminishing gains. The effect can be more pronounced in VER than in SER, since correcting a small number of residual symbol errors can recover an entire vector. For the 16-QAM configurations considered, we therefore use a short $J=3$ annealing schedule that captures most of the continuation gain while limiting the additional optimization cost.

Table~\ref{tab:smap_parameters} summarizes the default settings
used in the subsequent simulations. For coded simulations, the same parameters are used, and soft information is obtained from the
continuous SMAP estimate without $K_R$-neighbor refinement.

\begin{table}[t]
    \centering
    \caption{SMAP-KR parameters for MU-MIMO detection.}
    \label{tab:smap_parameters}
    \setlength{\tabcolsep}{5pt}
    \renewcommand{\arraystretch}{1.05}
    \begin{tabular}{@{}lccc@{}}
        \toprule
        Modulation & Stages $J$
        & $\bar \tau^2$ & $K_{\mathrm R}$ \\
        \midrule
        QPSK   & 1 & $0.10$                     & 5 \\
        16-QAM & 3 & $0.12 \to 0.10 \to 0.08$ & 5 \\
        \bottomrule
    \end{tabular}
\end{table}

\subsection{Uncoded MU-MIMO Detection}
\label{subsec:uncoded_results}

We next compare SMAP-KR with the baseline detectors in critically loaded MU-MIMO systems for QPSK and 16-QAM.

\subsubsection{QPSK}

\begin{figure*}[t!]
    \centering
    \begin{minipage}{0.32\textwidth}
        \centering
        \includegraphics[width=\linewidth,keepaspectratio]
        {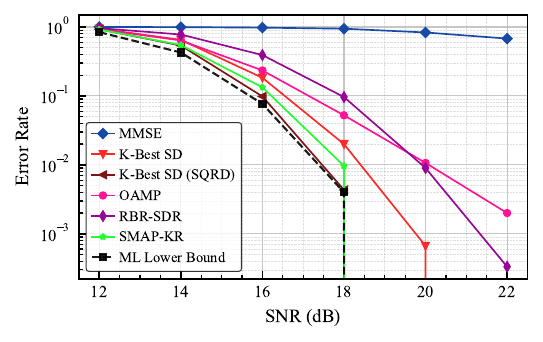}
        \textbf{(a)} $n=60$
    \end{minipage}
    \hfill
    \begin{minipage}{0.32\textwidth}
        \centering
        \includegraphics[width=\linewidth,keepaspectratio]
        {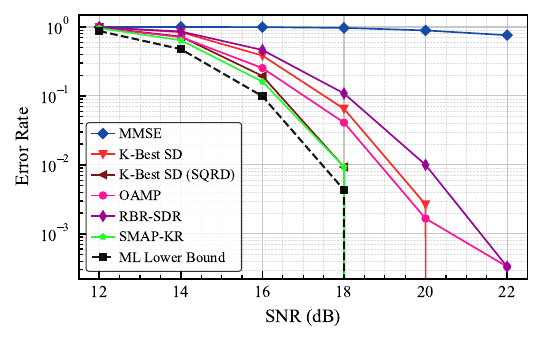}
        \textbf{(b)} $n=80$
    \end{minipage}
    \hfill
    \begin{minipage}{0.32\textwidth}
        \centering
        \includegraphics[width=\linewidth,keepaspectratio]
        {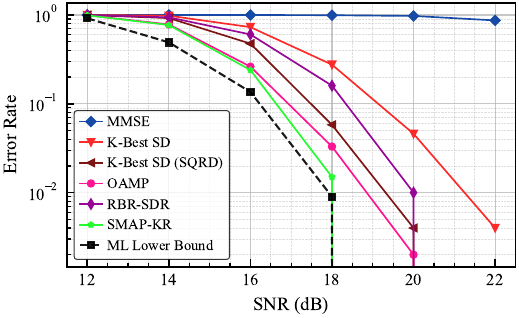}
        \textbf{(c)} $n=120$
    \end{minipage}
    \caption{Uncoded VER performance for QPSK ($|\mathcal{X}|=2$).}
    \label{fig:SER_QPSK}
\end{figure*}

Figure~\ref{fig:SER_QPSK} reports the QPSK results for $n=60$, $80$, and $120$. At $n=60$, SMAP-KR performs comparably to K-best SQRD, but its advantage increases with dimension. At $n=80$, it outperforms K-best SQRD, while at $n=120$, it achieves the best performance among the practical detectors considered. In contrast, the fixed-width tree search becomes increasingly restrictive as the dimension grows, while OAMP becomes competitive with K-best SQRD at the largest dimension.

The gap between SMAP-KR and the empirical ML lower bound remains approximately constant as the dimension increases. At $n=120$ and a target VER of $10^{-2}$, SMAP-KR is approximately $0.5$~dB from the bound, compared with $1.4$~dB for K-best SQRD and $0.9$~dB for OAMP. Thus, SMAP-KR retains a relatively small gap to the ML benchmark as the detection dimension increases.

\subsubsection{16-QAM}

\begin{figure*}[t]
    \centering
    \begin{minipage}{0.32\textwidth}
        \centering
        \includegraphics[width=\linewidth,keepaspectratio]
        {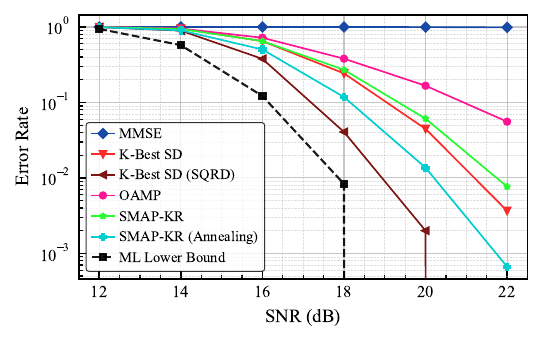}
        \textbf{(a)} $n=60$
    \end{minipage}
    \hfill
    \begin{minipage}{0.32\textwidth}
        \centering
        \includegraphics[width=\linewidth,keepaspectratio]
        {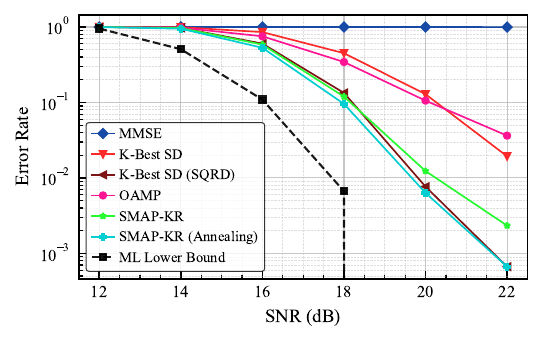}
        \textbf{(b)} $n=80$
    \end{minipage}
    \hfill
    \begin{minipage}{0.32\textwidth}
        \centering
        \includegraphics[width=\linewidth,keepaspectratio]
        {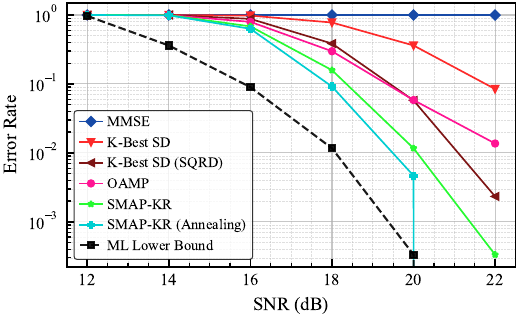}
        \textbf{(c)} $n=120$
    \end{minipage}
    \caption{Uncoded VER performance for 16-QAM ($|\mathcal{X}|=4$).}
    \label{fig:SER_16QAM}
\end{figure*}

\begin{figure*}[t!]
    \centering
    \begin{minipage}{0.32\textwidth}
        \centering
        \includegraphics[width=\linewidth,keepaspectratio]
        {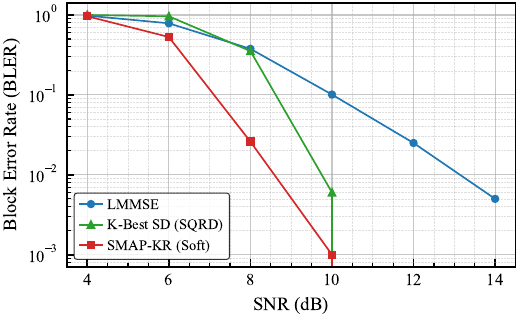}
        \textbf{(a)} $n=80$, QPSK
    \end{minipage}
    \hfill
    \begin{minipage}{0.32\textwidth}
        \centering
        \includegraphics[width=\linewidth,keepaspectratio]
        {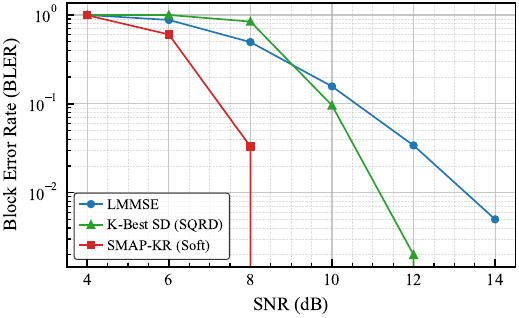}
        \textbf{(b)} $n=120$, QPSK
    \end{minipage}
    \hfill
    \begin{minipage}{0.32\textwidth}
        \centering
        \includegraphics[width=\linewidth,keepaspectratio]
        {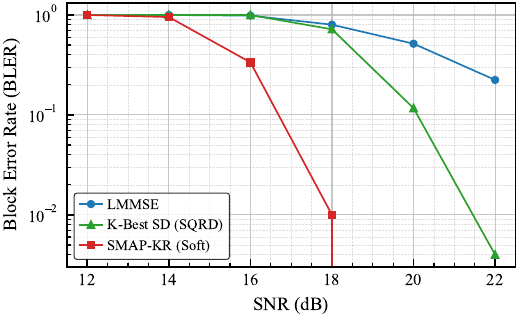}
        \textbf{(c)} $n=80$, 16-QAM
    \end{minipage}
    \caption{Coded BLER performance with a rate-$1/2$ LDPC code.}
    \label{fig:BLER}
\end{figure*}

Figure~\ref{fig:SER_16QAM} presents the corresponding 16-QAM results, using annealed SMAP-KR. SDR-RBR is omitted because the considered implementation is restricted to binary alphabets. At $n=60$, K-best SQRD provides the best performance, while at $n=80$ annealed SMAP-KR slightly outperforms it. At $n=120$, the advantage becomes substantial, with approximately a $1.4$~dB gain over K-best SQRD at a target VER of $10^{-2}$.

The gap to the empirical ML lower bound is larger than for QPSK, reflecting the more challenging optimization associated with the higher-order alphabet. Nevertheless, annealing allows SMAP-KR to improve relative to the competing detectors as the detection dimension increases.

\subsection{Coded Performance}
\label{subsec:coded_perf}

We finally evaluate the soft-output SMAP detector of Section~\ref{sec:soft_smap}. The coded simulations employ a rate-$1/2$ LDPC code with belief-propagation decoding and a maximum of 60 iterations. Figure~\ref{fig:BLER} reports the resulting BLER for QPSK and 16-QAM.

For comparison, LMMSE derives its soft information from a linear estimator, whereas soft K-best obtains approximate LLRs from the candidate vectors representing the competing bit hypotheses \cite{LSD,SSD}. The latter can overestimate reliability when a good counterhypothesis is absent from the finite candidate list. Soft SMAP instead extracts reliability from the local curvature of the constellation-aware SMAP objective without requiring explicit counterhypothesis candidates.

As shown in Fig.~\ref{fig:BLER}, soft SMAP provides a substantial BLER gain over both LMMSE and soft K-best in all considered configurations.  The gain persists for both QPSK and 16-QAM and at different system dimensions, indicating that the local Gaussian approximation provides useful reliability information despite representing only the neighborhood of the selected SMAP solution. 

These results also demonstrate that the same deterministic SMAP formulation can provide both hard estimates and soft reliability information. The latter requires primarily the additional $\mathcal{O}(n^3)$ curvature factorization, without posterior sampling or enlargement of a discrete candidate list.

\subsection{System Load}
\label{subsec:system_load}

\begin{figure}[t]
    \centering
    \includegraphics[width=0.7\linewidth,keepaspectratio]
    {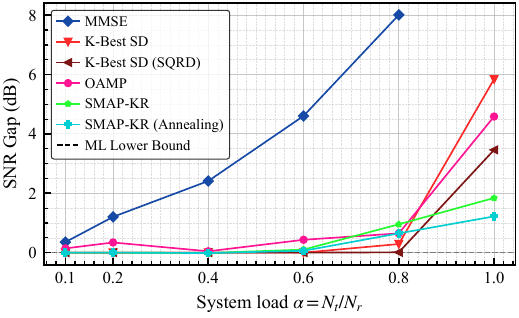}
    \caption{SNR gap (dB) between detectors and estimated ML VER lower bound for target VER = $10^{-2}$ ($n=120$, $|\mathcal{X}|=4$).}
    \label{fig:ver_alpha}
\end{figure}

To validate our method across different scenarios, we evaluate performance under varying system loads $\alpha = N_t/N_r$. Fig.~\ref{fig:ver_alpha} shows that higher values of $\alpha$ make the detection increasingly difficult. For small system loads, the gap is similar across all detectors, demonstrating that even low-complexity detectors can achieve satisfactory performance. As $\alpha$ increases, SMAP-KR remains competitive with search methods, outperforming the K-Best SQRD in the critically loaded scenario.

\subsection{Empirical Computational Cost}
\label{subsec:empirical_complexity}

We conclude by examining the computational behavior of SMAP-KR, complementing the analytical costs derived in Sections~\ref{sec:smap} and~\ref{sec:soft_smap}. In particular, the cost of continuous SMAP depends on the data-dependent number of objective and gradient evaluations required by L-BFGS-B.

Table~\ref{tab:smap_optimizer_it} reports the average number of gradient evaluations, $T_{\nabla}$, at the lowest and highest simulated SNRs.

\begin{table}[t!]
\centering
\setlength{\tabcolsep}{5pt}
\begin{tabular}{c c cc}
\toprule
\textbf{$n$}
& \textbf{Mod.}
& $T_{\nabla}$ at $\mathrm{SNR}_{\min}$
& $T_{\nabla}$ at $\mathrm{SNR}_{\max}$ \\
\midrule
60  & QPSK   & 28 & 21 \\
60  & 16-QAM & 40 & 32 \\
\midrule
80  & QPSK   & 31 & 22 \\
80  & 16-QAM & 45 & 33 \\
\midrule
120 & QPSK   & 35 & 22 \\
120 & 16-QAM & 53 & 34 \\
\bottomrule
\end{tabular}
\caption{Average number of gradient evaluations $T_{\nabla}$ at the lowest and highest simulated SNRs.}
\label{tab:smap_optimizer_it}
\end{table}

For QPSK, $T_{\nabla}$ increases only moderately with dimension at low SNR and remains nearly constant at high SNR. The 16-QAM case requires more evaluations, consistent with its more challenging multimodal objective. In both cases, fewer evaluations are required at high SNR, where the initialization is typically more accurate. These measurements characterize the empirical factor multiplying the $\mathcal{O}(n^2)$ per-evaluation cost in~\eqref{eq:smap_optimization_complexity_TN}; $T_{\nabla}$ should not be compared directly with a tree-search candidate width as an operation count.

Table~\ref{tab:runtime_01} provides an implementation-level comparison for the binary real-valued alphabet.

\begin{table}[t!]
\centering
\setlength{\tabcolsep}{3pt}
\begin{tabular}{l c c c}
\toprule
\textbf{Method}
& \textbf{$n=60$}
& \textbf{$n=80$}
& \textbf{$n=120$} \\
\midrule
LMMSE             & $0.20 \pm 0.00$ & $0.25 \pm 0.00$ & $0.54 \pm 0.00$ \\
OAMP              & $4.2 \pm 0.4$   & $5.5 \pm 0.6$   & $8.9 \pm 0.6$ \\
SDR-RBR           & $41.3 \pm 0.8$  & $54.9 \pm 1.6$  & $99.6 \pm 2.7$ \\
K-best SD         & $3.5 \pm 0.1$   & $4.4 \pm 0.1$   & $12.6 \pm 0.4$ \\
K-best SD (SQRD)  & $6.7 \pm 0.1$   & $8.8 \pm 0.25$  & $16.9 \pm 0.5$ \\
SMAP-KR           & $7.9 \pm 0.7$   & $8.7 \pm 1.2$   & $12.1 \pm 1.6$ \\
\bottomrule
\end{tabular}
\caption{Average runtime and standard deviation in milliseconds for the binary real-valued alphabet ($|\mathcal{X}|=2$).}
\label{tab:runtime_01}
\end{table}

LMMSE has the lowest runtime, and OAMP is also faster than SMAP-KR, whereas SDR-RBR is substantially more expensive. Relative to K-best SQRD, SMAP-KR requires $7.9$ versus $6.7$~ms at $n=60$, essentially the same runtime at $n=80$, and $12.1$ versus $16.9$~ms at $n=120$.  This behavior reflects the different algorithmic structures: with the two-neighbor approximation, SMAP is dominated by repeated $\mathcal{O}(n^2)$ dense operations followed by a small $K_R$-neighbor search, whereas fixed-width tree search operates over an increasingly deep detection tree.

Together with Figs.~\ref{fig:SER_QPSK} and~\ref{fig:SER_16QAM}, these measurements show an increasingly favorable performance--runtime tradeoff for SMAP-KR as the dimension grows. At $n=120$, SMAP-KR provides both better detection performance and lower measured runtime than the evaluated K-best SQRD implementation. Absolute runtimes remain implementation-dependent and are reported to complement rather than replace the analytical complexity results.

\section{Conclusion}
\label{sec:conclusion}

We proposed SMAP, a deterministic smoothed-MAP framework for high-dimensional critically loaded MU-MIMO detection. By replacing the discrete constellation prior with a Gaussian-mixture approximation, SMAP transforms discrete MAP detection into a differentiable continuous optimization problem while preserving the constellation structure. The resulting formulation can operate standalone or refine the output of another detector. For hard detection, SMAP-KR combines the continuous solution with a local $K_R$-neighbor search using the original ML metric, while annealing and a two-neighbor prior approximation improve robustness and efficiency for higher-order constellations.

We also developed a soft-output extension that extracts uncertainty from the local curvature of the same SMAP objective and converts it into bit LLRs. Unlike approaches for which soft-output generation requires additional candidate lists, counterhypothesis searches, or sampling procedures, SMAP extends naturally to soft detection through the local geometry of its smoothed posterior. Numerical results show that SMAP-KR becomes increasingly effective relative to the considered baselines as the detection dimension grows, providing substantial gains over LMMSE and K-best detection and a favorable performance--runtime tradeoff at the largest dimensions considered. Soft SMAP likewise provides substantial coded BLER gains over LMMSE- and K-best-based soft detection.

Several extensions are of interest for future work. In particular, the continuous SMAP formulation is applicable beyond the critically loaded setting considered here, motivating its investigation for overloaded MIMO systems. Adaptive selection of the smoothing and annealing parameters could further improve the performance--complexity tradeoff. For soft detection, scalable approximations to the inverse curvature diagonal could reduce the cubic cost of computing marginal uncertainties in very large systems. Channels with nonlinearities~\cite{CY18,safa2024data} can also be tackled with the same approach. 

\bibliographystyle{IEEEtran}
\bibliography{bib}

@article{safa2024data,
  title={Data detection in 1-bit quantized {MIMO} systems},
  author={Safa, Khodor and Combes, Richard and de Lacerda, Raul and Yang, Sheng},
  journal = {IEEE Trans. Commun.},
  volume={72},
  number={9},
  pages={5396--5410},
  year={2024},
  publisher={IEEE}
}

@ARTICLE{MaMIMO,
  author={Larsson, Erik G. and Edfors, Ove and Tufvesson, Fredrik and Marzetta, Thomas L.},
  journal={IEEE Commun. Mag.}, 
  title={Massive {MIMO} for next generation wireless systems}, 
  year={2014},
  volume={52},
  number={2},
  pages={186-195},
  doi={10.1109/MCOM.2014.6736761}}

@ARTICLE{MaMIMO_survey,
  author={Albreem, Mahmoud A. and Juntti, Markku and Shahabuddin, Shahriar},
  journal={IEEE Commun. Surveys Tuts.}, 
  title={Massive {MIMO} detection techniques: A Survey}, 
  year={2019},
  volume={21},
  number={4},
  pages={3109-3132},
  doi={10.1109/COMST.2019.2935810}}

@ARTICLE{50years,
  author={Yang, Shaoshi and Hanzo, Lajos},
  journal={IEEE Commun. Surveys Tuts.}, 
  title={Fifty years of {MIMO} detection: The road to large-scale {MIMOs}}, 
  year={2015},
  volume={17},
  number={4},
  pages={1941-1988},
  doi={10.1109/COMST.2015.2475242}}

@INPROCEEDINGS{ILS_comp,
  author={Hassibi, Babak and Vikalo, Haris},
  booktitle={Proc. IEEE Int. Conf. Acoust., Speech Signal Process. (ICASSP)},  
  title={{On the expected complexity of integer least-squares problems}}, 
  year={2002},
  volume={2},
  number={},
  pages={II-1497--II-1500},
  doi={10.1109/ICASSP.2002.5744897}}

@ARTICLE{MLD,
  author={Damen, M.O. and El Gamal, H. and Caire, G.},
  journal={IEEE Trans. Inf. Theory}, 
  title={{On maximum-likelihood detection and the search for the closest lattice point}}, 
  year={2003},
  volume={49},
  number={10},
  pages={2389-2402},
  doi={10.1109/TIT.2003.817444}}

@ARTICLE{K_best,
  author={Guo, Zhan and Nilsson, P.},
  journal={IEEE J. Sel. Areas Commun.}, 
  title={{Algorithm and implementation of the {K}-best sphere decoding for {MIMO} detection}}, 
  year={2006},
  volume={24},
  number={3},
  pages={491-503},
  doi={10.1109/JSAC.2005.862402}}

@INPROCEEDINGS{IFSD,
  author={Barbero, L.G. and Thompson, J.S.},
  booktitle={Proc. IEEE Int. Conf. Acoust., Speech Signal Process. (ICASSP)}, 
  title={{Performance analysis of a fixed-complexity sphere decoder in high-dimensional MIMO Systems}}, 
  year={2006},
  volume={4},
  number={},
  pages={IV-557--IV-560},
  doi={10.1109/ICASSP.2006.1661029}}

@INPROCEEDINGS{Yao_Wornell_LR,
  title={Lattice-reduction-aided detectors for {MIMO} communication systems},
  author={Yao, Huan and Wornell, Gregory W},
  booktitle={Proc. IEEE Global Telecommun. Conf. (GLOBECOM)},
  volume={1},
  pages={424--428},
  year={2002},
  doi={10.1109/GLOCOM.2002.1188114}}

@inproceedings{windpassinger2003low,
  title={Low-complexity near-maximum-likelihood detection and precoding for {MIMO} systems using lattice reduction},
  author={Windpassinger, Christoph and Fischer, Robert F. H.},
  booktitle={Proc. IEEE Inf. Theory Workshop (ITW)},
  pages={345--348},
  year={2003},
  doi={10.1109/ITW.2003.1216764}}

@article{Hayakawa_LR_overloaded,
  author  = {Ryo Hayakawa and Kazunori Hayashi and Megumi Kaneko},
  title   = {Lattice reduction-aided detection for overloaded {MIMO} using slab decoding},
  journal = {IEICE Trans. Commun.},
  year    = {2016},
  volume  = {E99-B},
  number  = {8},
  pages   = {1697--1705},
  doi     = {10.1587/transcom.2015CCP0014}
}

@ARTICLE{GenPGD,
  author={He, Lanxin and Wang, Zheng and Yang, Shaoshi and Liu, Tao and Huang, Yongming},
  journal={IEEE Trans. Wireless Commun.}, 
  title={{Generalizing projected gradient descent for deep-learning-aided massive MIMO detection}}, 
  year={2024},
  volume={23},
  number={3},
  pages={1827-1839},
  doi={10.1109/TWC.2023.3292124}}

@INPROCEEDINGS{PGD_net,
  author={Takabe, Satoshi and Imanishi, Masayuki and Wadayama, Tadashi and Hayashi, Kazunori},
  booktitle={Proc. IEEE Int. Conf. Commun. (ICC)}, 
  title={{Deep learning-aided projected gradient detector for massive overloaded MIMO channels}}, 
  year={2019},
  volume={},
  number={},
  pages={1-6},
  doi={10.1109/ICC.2019.8761049}}

@INPROCEEDINGS{ADMMNet,
  author={Un, Man-Wai and Shao, Mingjie and Ma, Wing-Kin and Ching, P. C.},
  booktitle= {Proc. IEEE Data Sci. Workshop (DSW)},
  title={Deep {MIMO} detection using {ADMM} unfolding}, 
  year={2019},
  volume={},
  number={},
  pages={333-337},
  doi={10.1109/DSW.2019.8755566}}

@ARTICLE{OAMP,
  author={Ma, Junjie and Ping, Li},
  journal={IEEE Access}, 
  title={{Orthogonal AMP}}, 
  year={2017},
  volume={5},
  number={},
  pages={2020-2033},
  doi={10.1109/ACCESS.2017.2653119}}

@INPROCEEDINGS{OAMP_MA,
  author={Jeon, Charles and Ghods, Ramina and Maleki, Arian and Studer, Christoph},
  booktitle={Proc. IEEE Int. Symp. Inf. Theory (ISIT)}, 
  title={Optimality of large {MIMO} detection via approximate message passing}, 
  year={2015},
  volume={},
  number={},
  pages={1227-1231},
  doi={10.1109/ISIT.2015.7282651}}

@ARTICLE{CMDNet,
  author={Beck, Edgar and Bockelmann, Carsten and Dekorsy, Armin},
  journal={IEEE Trans. Commun.}, 
  title={{CMDNet}: Learning a probabilistic relaxation of discrete variables for soft detection with low complexity}, 
  year={2021},
  volume={69},
  number={12},
  pages={8214-8227},
  doi={10.1109/TCOMM.2021.3114682}}

@INPROCEEDINGS{SOAV_MAP,
  author={Hayakawa, Ryo and Hayashi, Kazunori},
  booktitle={Proc. Int. Symp. Inf. Theory Appl. (ISITA)}, 
  title={Error recovery with relaxed {MAP} estimation for massive {MIMO} signal detection}, 
  year={2016},
  volume={},
  number={},
  pages={478-482},
  doi={}}

@ARTICLE{HOMTD,
  author={Shao, Mingjie and Ma, Wing-Kin and Liu, Junbin},
  journal={IEEE Open J. Signal Process.}, 
  title={An explanation of deep {MIMO} detection from a perspective of homotopy optimization}, 
  year={2023},
  volume={4},
  number={},
  pages={108-116},
  doi={10.1109/OJSP.2023.3243523}}

@ARTICLE{GD_XL_MIMO,
  author={Chen, Qiqiang and Wang, Zheng and Zhang, Chuan and Shu, Feng and Huang, Yongming and Niyato, Dusit},
  journal={IEEE Trans. Signal Process.}, 
  title={Distributed penalty {Nesterov} gradient algorithm for extremely large-scale {MIMO} systems}, 
  year={2026},
  volume={74},
  number={},
  pages={734-749},
  doi={10.1109/TSP.2026.3660267}}

@ARTICLE{Q_penalty,
  author={Chen, Qiqiang and Wang, Zheng and Qi, Chenhao and Shu, Feng and Huang, Yongming},
  journal={IEEE Trans. Commun.}, 
  title={Quantized penalty gradient algorithm for massive {MIMO} systems with low-resolution {ADCs}}, 
  year={2026},
  volume={74},
  number={},
  pages={3342-3355},
  doi={10.1109/TCOMM.2025.3646813}}

@ARTICLE{ALD,
  author={Zilberstein, Nicolas and Dick, Chris and Doost-Mohammady, Rahman and Sabharwal, Ashutosh and Segarra, Santiago},
  journal= {IEEE Trans. Wireless Commun.}, 
  title={Annealed {Langevin} dynamics for massive {MIMO} detection}, 
  year={2023},
  volume={22},
  number={6},
  pages={3762-3776},
  doi={10.1109/TWC.2022.3221057}}

@ARTICLE{SGLD,
  author={Wu, Zhiwen and Li, Hui},
  journal={IEEE Commun. Lett.}, 
  title={Stochastic gradient {Langevin} dynamics for massive {MIMO} detection}, 
  year={2022},
  volume={26},
  number={5},
  pages={1062-1065},
  doi={10.1109/LCOMM.2022.3151141}}

@INPROCEEDINGS{Tstu,
  author={Hagiwara, Junichiro and Matsumura, Kazushi and Asumi, Hiroki and Kasuga, Yukiko and Nishimura, Toshihiko and Sato, Takanori and Ogawa, Yasutaka and Ohgane, Takeo},
  booktitle={Proc. IEEE Global Commun. Conf. (GLOBECOM)}, 
  title={Near-optimal stochastic {MIMO} signal detection with a mixture of {t}-distributions prior}, 
  year={2023},
  volume={},
  number={},
  pages={6627-6633},
  doi={10.1109/GLOBECOM54140.2023.10437162}}

@ARTICLE{LBFGS,
  author={Liu, D.C. and Nocedal, J.},
  journal={Math. Program.}, 
  title={On the limited memory {BFGS} method for large scale optimization}, 
  year={1989},
  volume={45},
  pages={503-528},
  doi={/10.1007/BF01589116}}

@ARTICLE{LBFGS_B,
  author={Byrd, Richard H. and Lu, Peihuang and Nocedal, Jorge and Zhu, Ciyou},
  journal={SIAM J. Sci. Comput.},
  title={{A limited memory algorithm for bound constrained optimization}},
  year={1995},
  volume={16},
  number={5},
  pages={1190--1208},
  doi={10.1137/0916069}
}

@article{SQRD,
author = {W{\"u}bben, D. and B{\"o}hnke, R. and Rinas, J. and K{\"u}hn, V. and Kammeyer, K. D.},
title = {Efficient algorithm for decoding layered space-time codes},
journal ={Electron. Lett.},
volume = {37},
number = {22},
pages = {1348-1350},
year = {2001},
doi = {10.1049/el:20010899}}

@INPROCEEDINGS{SDR-RBR,
  author={Wai, Hoi-To and Ma, Wing-Kin and So, Anthony Man-Cho},
  booktitle={Proc. IEEE Int. Conf. Acoust., Speech Signal Process. (ICASSP)}, 
  title={Cheap semidefinite relaxation {MIMO} detection using row-by-row block coordinate descent}, 
  year={2011},
  volume={},
  number={},
  pages={3256-3259},
  doi={10.1109/ICASSP.2011.5946716}}

@ARTICLE{CY18,
  author={Combes, Richard and Yang, Sheng},
  journal={IEEE Trans. Commun.}, 
  title={An approximate {ML} detector for {MIMO} channels corrupted by phase noise}, 
  year={2018},
  volume={66},
  number={3},
  pages={1176--1189},
  doi={10.1109/TCOMM.2017.2774803}}

@ARTICLE{LSD,
  author={Hochwald, B.M. and ten Brink, S.},
  journal={IEEE Trans. Commun.}, 
  title={Achieving near-capacity on a multiple-antenna channel}, 
  year={2003},
  volume={51},
  number={3},
  pages={389--399},
  doi={10.1109/TCOMM.2003.809789}}

@INPROCEEDINGS{SSD,
  author={Studer, C. and Wenk, M. and Burg, A. and Bolcskei, H.},
  booktitle={Proc. 40th Asilomar Conf. Signals, Syst. Comput.}, 
  title={Soft-output sphere decoding: Performance and implementation aspects}, 
  year={2006},
  volume={},
  number={},
  pages={2071--2076},
  doi={10.1109/ACSSC.2006.355132}}

\end{document}